\documentclass[acmlarge,10pt,nonacm]{acmart}
\usepackage{ifthen}
\usepackage{fancyhdr}
\usepackage{xspace}
\usepackage{booktabs}
\usepackage{mathptmx}   % Times + Times-like math symbols
\usepackage{courier}
\usepackage{color}
\usepackage{graphicx}
\usepackage{multicol}
\usepackage{multirow}
\usepackage{adjustbox}
\usepackage{verbatim}
\usepackage{appendix}
\usepackage{titlesec}

\newcommand{\TITLE}{Hardware-Enforced Integrity and Provenance for Distributed Code Deployments}
\newcommand{\SUBTITLE}{Area: Guidelines for Software Integrity Chains and Provenance}

\newcommand{\ANONYMOUS}{no}
\newcommand{\SHOWTOAPPEAR}{no}
\newcommand{\COMMENTS}{yes}

\usepackage[absolute]{textpos}
\newcommand{\ToAppear}{%
\begin{textblock*}{\textwidth}(0.95in,0.4in)
\begin{flushright}
    \noindent{\fbox{\textsf{Draft version---please do not redistribute.}}}
\end{flushright}
\end{textblock*}
}

\usepackage{caption}
\usepackage{algorithm}
\usepackage[noend]{algpseudocode}

\usepackage{listings}

\newcommand{\eg}{e.g.,\xspace}
\newcommand{\ie}{i.e.,\xspace}

\newcommand{\Parabreak}{1.5ex}
\newcommand{\Paragraph}[1]{\vspace{\Parabreak}\noindent\textbf{#1}}
\newcommand{\swsupchain}{software supply chain\xspace}

\newcommand{\sgx}{Intel SGX\xspace}

\newcommand{\ignore}[1]{}

\ifthenelse{\equal{\COMMENTS}{yes}}{%
\newcommand{\msm}[1]{\textbf{MSM: #1}}
} {
\newcommand{\msm}[1]{}
}

\usepackage{enumitem}
\setlist[itemize]{noitemsep,nolistsep}
\setlist[enumerate]{noitemsep,nolistsep}
\usepackage{colortbl}
\definecolor{Gray}{gray}{0.9}
\definecolor{LightGreen}{rgb}{0.2,1,0.4}

\usepackage{longtable}

\date{}
\title{\TITLE}
\ifthenelse{\equal{\ANONYMOUS}{no}}
{
\author{Marcela S. Melara}
\affiliation{%
  \jobtitle{Research Scientist}
  \institution{Intel Labs}
  \city{Hillsboro}
  \state{OR}
  \country{USA}}
\email{marcela.melara@intel.com}

\author{Mic Bowman}
\affiliation{%
  \jobtitle{Senior Principal Engineer}
  \institution{Intel Labs}
  \city{Hillsboro}
  \state{OR}
  \country{USA}}
\email{mic.bowman@intel.com}
}{\author{}}
\subtitle{\SUBTITLE}

\begin{document}
\maketitle

\ifthenelse{\equal{\SHOWTOAPPEAR}{yes}}{\ToAppear}{}

\section{Introduction}
\label{sec:intro}
Distributed code deployments today rely very heavily on a complex series
of transformation and inspection operations, called the \emph{\swsupchain},
for the creation of an \emph{executable bundle} that is run at a cloud provider.
For example, a compilation \emph{tool} transforms one or more input
\emph{software artifacts} (\ie source code files and shared libraries)
generating one or more output artifacts (\ie bytecode or binary files, 
other shared libraries, container images). These may then subsequently be passed as input artifacts
to an automated testing tool, inspecting the functionality
of the artifacts without modifying them, and returning the test results.
Thus, we refer to a software artifact and attached metadata as an executable bundle
that can be deployed at a cloud provider.

However, a plethora of attacks have undermined the integrity 
of entire deployments by compromising one or more operations of the 
\swsupchain~\cite{solarwinds-fireeye,sgx-signing-injection,backstabbers-knife,left-pad-blog,left-pad-ars,poison-sw-supply,maloss,vulns-ci, sec-ci,cloud-assurance,off-my-cloud}.
Prior work~\cite{grafeas,kritis,binary-auth,in-toto} has proposed addressing these issues
by capturing metadata about software artifacts to obtain verifiable information about 
the supply chain of an application.
in-toto~\cite{in-toto}, in particular, realizes this approach relying on cryptographically
verifiable metadata about each individual supply chain operation to ensure
%By cryptographically linking the metadata for each operation, software developers may prior to deployment validate
%the integrity of the supply chain \emph{layout} end-to-end, \ie 
that an executable bundle was indeed created by the expected supply chain tools in the expected order.

Yet, as distributed computing continues to evolve, we observe three major trends that introduce
new security challenges to code deployment.

\textbf{1. Rise of the microservices.}
Rather than deploying a monolithic, self-contained, fully packaged application, 
developers increasingly decompose an application into small, independent
components.
These microservices are then uploaded at a cloud provider
enabling greater application portability, efficiency and fault isolation. 

\textbf{2. Shift towards lightweight containers.}
As cloud providers seek to optimize resource usage while improving microservice performance
they employ lighter weight execution environments that rely on 
software-based techniques for isolation in a multi-tenant setting (\eg~\cite{faasm,fastly-lucet}).
However, this means cloud providers can no longer rely on virtual machine-like
mechanisms to protect their resources against vulnerable or buggy 
code, and strongly isolate co-tenant microservices.

\textbf{3. On-demand deployment.}
To further optimize resource usage, hosting services determine on-demand 
which compute platforms are most suitable to deploy particular microservices.

\Paragraph{The core problem.}
These three trends demonstrate that today's distributed applications require 
more adaptable and comprehensive means to establish trust in deployed code.
That is, it is no longer sufficient to only ensure that code is generated by the
expected \swsupchain. 

Deployed microservices must adhere to a multitude of application-level security 
requirements and regulatory constraints imposed by mutually distrusting 
\emph{application principals}-- software developers, cloud providers, and even data owners. 
Although these principals wish to enforce their individual security requirements,
they do not currently have a common way of easily identifying, expressing and automatically 
enforcing these requirements at deployment time.

\Paragraph{Our Proposal.}
CDI (Code Deployment Integrity) is a security policy framework 
that enables distributed application principals to establish trust 
in deployed code through \emph{high-integrity} provenance information.

We observe that principals expect the \swsupchain to preserve
certain code security properties throughout the creation of an executable bundle, 
even if the code is transformed or inspected through various tools
(\eg compilation inserts stack canaries for memory safety).
Our key insight in designing CDI is that even if application principals do not trust each other directly,
they can trust a microservice bundle to meet their security policies
\emph{if they can trust the tools involved in creating the bundle}.

%Crucially, CDI does \emph{not} directly evaluate or guarantee the correctness or security
%of microservice code. Instead, the provenance information obtained
%throughout the \swsupchain provides an orchestrator with the 
%\emph{history of transformations and inspections} performed to create a given 
%microservice. Because of the code properties associated
%with specific supply chain operations, as long as the orchestrator can establish trust in 
%the provenance of a microservice, it can gain assurances about the security properties of the code by
%proxy of its history. 
 
%Though our approach of CDI is general, we demonstrate our approach with 
%a proof-of-concept based on the Intel SGX~\cite{sgx,sgx-paper} TEE and the Private Data 
%Objects~\cite{pdo-git} confidential smart contract framework,
%which deploys code only after verifying its provenance.
\section{Code Deployment Integrity}
As a microservice is created through a \swsupchain,
each tool in the supply chain digitally signs its input and
output as well as operation-specific metadata 
(\eg compiler flags, testing configuration).
Crucially, CDI metadata also embeds a cryptographic hash of the metadata
attached to each input artifact generated by ``upstream'' operations forming a hash tree 
that represents an auditable \emph{provenance chain} for the entire supply chain
of the executable microservice bundle received by the cloud provider.

\subsection{Key Components}
Protecting the integrity of the CDI metadata and efficiently auditing 
the provenance chain is vital to the security and adoptability of CDI.
We address these important challenges with two key mechanisms.

\textbf{1. Trusted execution environments (TEEs).}
CDI leverages TEEs (\eg~\cite{sgx,sgx-paper,amd-sev,tdx}) for their 
\emph{hardware-enforced} integrity and authentication properties for code and data.
Specifically, running individual \swsupchain tools in a TEE such as 
\sgx~\cite{sgx} can help CDI extend and enforce provenance in hardware. 
As tools derive their signing keys within an enclave, the digitally signed CDI metadata 
is bound to the specific platform that generated the keys.
%In addition, TEEs may help preserve the confidentiality of privacy-sensitive data 
%during supply chain operations (\eg proprietary source code at build time).

\textbf{2. Vetting authorities.}
CDI introduces independent entities called vetting authorities
that are responsible for certifying that specific \swsupchain tools preserve the 
expected security properties.
Specifically, CDI requires vetting authorities to produce digitally signed tool certifications
that can be embedded in a tool's provenance metadata.
Additionally, vetting authorities could leverage TEE technology 
to remotely authenticate the TEE-enabled tools they are certifying. 

In practice, we envision each vetting authority will determine the exact process
by which to vet individual supply chain tools. 
To enable vetting a large array of supply chain tools available to developers and
cloud providers today, CDI supports a hierarchy of vetting authorities
akin to how traditional certificate authorities operate today.

\subsection{Design Goals}
Our design of CDI meets the following three goals.

\textbf{G1: Code provenance can be validated without revealing 
any software artifacts.}
Only the cryptographic hashes (\eg SHA-256)
of both input and output artifacts are included in CDI metadata.
Thus, application principals need not share proprietary or highly privacy-sensitive
artifacts directly with other principals to enable validation.

\textbf{G2: Application principals need not rely on a single centralized
root of trust.}  %For instance, a given vetting authority may specialize in formal 
% verification of C/C++ compilation toolchains.
Developers, cloud providers and data owners may independently select the vetting
authorities they trust to properly certify tools in the \swsupchain ecosystem.
Further, CDI allows principals to require a threshold number of trusted vetting authorities
to have certified a given tool for added confidence in its operation.
At validation time, CDI ensures that a microservice's provenance meets all 
specified trust policies.

\textbf{G3: Trust in code can be established at deployment time
without a priori knowledge of the tools that created the code.}
As long as a chain of trust between a vetting authority and a given tool can 
be established, the CDI provenance metadata is sufficient
to gain trust in a microservice and ensure it enforces a principal's security 
policy.
\section{Conclusion}
We have presented CDI (Code Deployment Integrity),
a framework for verifiably capturing and validating microservice security properties 
and requirements via high-integrity metadata about the \swsupchain.
Our goal is to enable security-oriented code deployment,
in which provenance metadata can be used to enforce complex security policies imposed by
a multitude of application principals.
By leveraging trusted execution environments and vetting
authorities that act as trusted intermediaries, CDI provides strong integrity for 
code provenance metadata while reducing adoption efforts for software developers 
compared to prior approaches.

%\section*{Acknowledgments}

\bibliographystyle{ACM-Reference-Format}
\bibliography{references}

%\clearpage
%\onecolumn
%\appendix
%\input{appendix}

\end{document}